\documentclass[acmsmall, screen]{acmart}

\AtBeginDocument{%
  }

\setcopyright{none}
\copyrightyear{2023}
\acmYear{2023}
\acmDOI{XXXXXXX.XXXXXXX}

\acmVolume{0}
\acmNumber{0}
\acmArticle{0}
\acmMonth{0}

\begin{document}

\title{What's my role? Modelling responsibility for AI-based safety-critical systems}

\author{Philippa Ryan}
\orcid{0000-0003-1307-5207}
\author{Zoe Porter}
\orcid{0000-0002-4467-3288}
\author{John McDermid}
\orcid{0000-0003-4745-4272}
\author{Ibrahim Habli}
\orcid{0000-0003-2736-8238}
%\author[inst1,inst2]{Author Three}
\affiliation{\institution{Department Of Computer Science, University of York}%Department and Organization
            \city{York}
            \postcode{YO10 5DD}
            \country{UK}}
\author{Joanna Al-Qaddoumi}
\orcid{0009-0005-6340-9983}
\affiliation{  \institution{York Law School, University of York}
  \city{York}
  \postcode{YO10 5GD}
  \country{UK}}

\renewcommand{\shortauthors}{Ryan et al.}

%TC:ignore
\begin{abstract}
AI-Based Safety-Critical Systems (AI-SCS) are being increasingly developed and deployed in the real world. These can pose a risk of harm to people and the environment, hence reducing that risk is an overarching priority during development and operation. Many contain Machine Learning (ML) components whose performance is hard to predict. As more AI systems become autonomous, a layer of risk management via human intervention has been removed. 
Following an accident it will be important to identify causal contributions and the different responsible actors behind those to learn from mistakes and prevent similar future events. Many authors have commented on the "responsibility gap" where it is difficult for developers and manufacturers to be held responsible for behaviour of an AI-SCS which contributes to harm. This is due to the complex development cycle for AI, the uncertainty in black-box AI components performance, and dynamic operating environment. Instead, a human operator of the AI-SCS can become a "liability sink" absorbing blame for the consequences of AI-SCS outputs they weren’t responsible for creating, and may not have sufficient understanding of.

This cross-disciplinary paper considers different senses of responsibility (role, moral, legal and causal), and how they apply in the context of AI-SCS safety. We use a core conceptual formulation \textit{Actor(A) is responsible for Occurrence(O)} to create detailed role responsibility models, including related tasks and resources. Our aim is to present a practical method to precisely capture responsibility relationships, and provide clarity on the previously identified responsibility issues. We propose an analysis method to review the models for safety impacts. Our paper demonstrates the utility of the approach with two practical examples. The first is a retrospective analysis of the Tempe Arizona fatal collision involving an autonomous vehicle. The second is a safety focused predictive role-responsibility analysis for an AI-based diabetes co-morbidity prediction system. We show how our notation and analysis can illuminate responsibility issues during the AI development process and identify the impact of uncertainty surrounding how tasks were performed. For both examples, our focus was primarily on safety, with an aim of reducing unfair or disproportionate blame being placed on operators or developers of AI-SCS. We present a discussion and avenues for future research, including the development of a richer causal contribution model.

%We have previously published an analysis of the Tempe Arizona fatal collision in \cite{TAS_symp}, but this did not include the modelling notation and analysis method presented here and did not include the legal conclusion. We have also previously published work analysing safety impact of training data issues on the AI-based diabetes co-morbidity prediction system, but this was not related to role responsibility \cite{DCP_safecomp}. The core responsibility formulation can be found in \cite{unravelling_resp}, but this paper adds a novel responsibility notation and analysis to that work.
\end{abstract}
%TC:endignore


%TC:ignore
\begin{CCSXML}
<ccs2012>
<concept>
<concept_id>10010583.10010750.10010769</concept_id>
<concept_desc>Hardware~Safety critical systems</concept_desc>
<concept_significance>500</concept_significance>
</concept>
<concept>
<concept_id>10010520.10010575</concept_id>
<concept_desc>Computer systems organization~Dependable and fault-tolerant systems and networks</concept_desc>
<concept_significance>500</concept_significance>
</concept>
<concept>
<concept_id>10011007.10010940.10011003.10011114</concept_id>
<concept_desc>Software and its engineering~Software safety</concept_desc>
<concept_significance>500</concept_significance>
</concept>
</ccs2012>
\end{CCSXML}


\ccsdesc[500]{Hardware~Safety critical systems}
\ccsdesc[500]{Software and its engineering~Software safety}
\ccsdesc[500]{Computer systems organization~Dependable and fault-tolerant systems and networks}

\keywords{AI-based systems, Safety, Responsibility}

\received{December 2023}


\maketitle
%TC:endignore

\section{Introduction}

AI-Based Safety-Critical Systems (AI-SCS), such as autonomous vehicles, inspection drones, medical diagnosis systems and others, are increasingly being developed and deployed in the real world. AI-based systems provide many technical and societal challenges, particularly when they include black box Machine Learning (ML) components \cite{ashmore2021assuring} and are used in a dynamic operating environment. AI-SCS typically have a complex development chain with many actors, including software developers, data scientists, regulators, service providers, suppliers \cite{many_hands_ai, unravelling_resp}, whose roles all contribute to safety. This provides many areas of uncertainty, and difficulty in assuring they will behave in an acceptably safe way through-life. Additionally, for many AI-SCS there is limited or no interaction with a human operator \cite{sujan2023looking, sujan2019human} and there are gaps understanding who was responsible and/or liable following an incident or accident \cite{burton2020mind}. Human operators who do compensate for uncertainty in AI behaviour, face an unfair burden of risk, absorbing blame for the consequences of system outputs they weren’t responsible for creating \cite{lawton2023clinicians}. 

The notion of responsibility has long been studied in a philosophy and law, often with a backwards-looking perspective to understand events that have occurred. However, it has not been studied in depth in safety engineering, which also requires a forwards-looking perspective to predict potential harm and prevent it or reduce its severity. Nor has it been practically applied in the context of AI-SCS, although there is much research into responsibility issues for AI-SCS. This paper considers how a number of different senses of responsibility (role, moral, legal and causal \cite{hart}) apply to actors developing and operating AI-SCS. We have an emphasis on role responsibility, and demonstrate how analysis can help to uncover gaps, risk transfer, and burden of risk when developing and operating AI-SCS.

This paper provides the following contributions. First, we extend earlier work on responsibility modelling \cite{unravelling_resp} and task analysis \cite{resp_model_sommer,resp_model_Lock} to develop a notation and method for modelling role responsibility and role relationships for AI-SCS. This can be used to describe and link actors, occurrences and resources across AI-SCS lifecycle phases. Second, we apply the notation to two different AI-SCS scenarios, showing how our models illuminate issues (e.g., gaps, conflicts) of responsibility and provide greater clarity on risk distribution and safety impact. We particularly focus on how engineering roles and technical challenges during AI development can lead to burden of risk for operational roles.

In section \ref{sec:background} we review related work in this area, considering problems about the attribution of responsibility for AI-SCS, and motivate the need for responsibility analysis. In section \ref{sec:method} we present our responsibility modelling notation, and a method for analysing and reviewing role responsibility models. 
Section \ref{sec:uber_case_study} contains our first example study of responsibility modelling, using the accident findings of the fatal collision of a pedestrian with an autonomous vehicle in Tempe, Arizona. Section \ref{sec:diabetes_case_study} contains a second example study of an AI-based diabetes co-morbidity predictor, used to support a clinician during medical diagnoses of Type II diabetes patients. We present a complex forwards-looking model covering both development of the predictor and its operational use. We show how this model can be analysed to determine role responsibility issues. Finally, there is a discussion and considerations for future work in section \ref{sec:conc}.

\section{Background and related literature}
\label{sec:background}
\subsection{AI-based Safety critical systems and responsibility gaps}

Safety-critical systems are defined as those whose failure, under certain conditions, can lead to harm to humans or the environment \cite{61508}. There are many different examples, such as medical devices, nuclear power plants, defence systems, cars and aircraft. AI-SCS include AI elements to replace some or all of a human function, such as driverless autonomous vehicles, unmanned inspection drones, and medical diagnoses. An AI-SCS is subject to differing legal and regulatory regimes, depending on domain, type, country of use and even degree of autonomy. Rather than attempt to cover all these we briefly summarise some common practices which will broadly apply when developing an AI-SCS.

Typically the manufacturers and operators of an AI-SCS will provide a Safety Assurance Case (SAC) or safety justification that the system is \textit{acceptably safe} \cite{kelly1998arguing, sujan2016should} for use in a defined context. This should demonstrate that all reasonable means of risk reduction have been enacted, by following established good engineering practice (e.g., software testing), adding operational controls (e.g., manual intervention), or reducing severity of accidents (e.g., sprinkler systems). Good practice is encapsulated in domain specific standards, such as \cite{14971} for medical devices, \cite{26262} for automotive electrical/electronic systems, \cite{178c} for avionics, and \cite{056} for defence systems (noting that consensus on good practice is not fully established for AI components \cite{ashmore2021assuring}). The SAC may require external and independent scrutiny, such as by a regulator. 
Hence, even though there is residual risk associated with operating the AI-SCS, it may not be considered fair to hold developer's and/or manufacturer's responsible for an accident if they exercised due diligence. The SAC may be used as evidence in such a situation (see Section \ref{sec:soft_law}).

There are established practices for modelling and analysing operational responsibility for tasks performed by humans for conventional critical systems (e.g., \cite{ONR_human_task, resp_model_sommer, resp_model_ORDIT, resp_model_Lock}). A complication for AI-SCS is that an AI component performs some or all of the decisions and tasks previously assigned to a human operator, including some to reduce the risk of accidents \cite{monkhouse2020enhanced, burton2020mind}. E.g., an autonomous drone will take actions to avoid collision, rather than a remote operator. Alternatively, an operator may be required to only make interventions when the AI-SCS cannot, increasing cognitive load \cite{BAINBRIDGE1983775, av_driver_takeover}. When there are accidents, the operator may absorb blame for the consequences of AI-SCS outputs they weren't responsible for creating, and may not have sufficient understanding of. This is termed "liability sink" \cite{lawton2023clinicians}. 

Many authors have commented on the "responsibility gap" for AI-SCS e.g. \cite{burton2020mind, Matthias_resp, gunkel2012machine_moral, morgan_book}, where it is difficult for developers and manufacturers to be held responsible for behaviour of an AI-SCS which contributes to harm. This is due to difficulty in a) ensuring safe behaviour of black-box AI component, and b) defining precise operating domains \cite{ashmore2021assuring}. It has even been suggested this uncertainty can be used to avoid responsibility (see \ref{sec:many_hands}). In \cite{resp_in_system}, the authors argue for the AI-based system itself to be programmed to be responsible, although practical methods to do this are unknown. 

Munch \cite{Munch2023}, notes that a responsibility gap could be viewed positively, because it avoids unfairly blaming individuals, and replacing human agency with AI can remove the burden of decision making. We reject the idea of not attempting to clarify and, if possible, reduce responsibility gaps. This is not to punish individuals, but to avoid putting people in the situation where they could face unfair blame\footnote{Two aspects of blame, moral and legal liability, are discussed later in the paper. We note that one can have liability without fault, and typically blame is only an appropriate response when someone is at fault.}, and so that we can learn from mistakes and avoid them in future. Good safety practice is based on a \textit{just culture} as proposed by \cite{dekker}\cite{justukgov}\cite{safety_culture_trans}, in which incident reporting is encouraged within an organisation, promoting transparency without fear of blame. In this atmosphere, taking responsibility can be encouraged. 

%\textcolor{red}{HERE}
The recent EU AI Act \cite{EU_AI_act} is intended to close some legal responsibility gaps. The current draft has been criticised for not considering some of the unique problems of AI, including its dynamic nature, its complex lifecycle, understanding end user rights \cite{EdwardsAIAct}, and not tracing responsibility back to AI developers \cite{eu_act_hrw}. The EU AI liability directive \cite{EU_AI_liability} is intended to provide means of pursuing civil liability, but legal review suggests technical issues of identifying causal responsibility (responsibility gaps) are not addressed \cite{morgan_book}.

One way to ensure that the nearest human operator isn't held solely responsible for the consequences of AI-SCS behaviour after an accident is to articulate the roles and potential liability of all actors (e.g., safety engineers) with safety-related tasks more systematically earlier on in the system lifecycle. Further, this can protect actors, such as engineers or other decision makers, from unfair blame by demonstrating their tasks were performed with due diligence.

\subsection{Responsibility complexity in AI-SCS development and operation}
\label{sec:many_hands}

If we are to be clear about the responsibility of actors during the development of AI-SCS, we need understand the complex supply, development and operation chain for AI \cite{EdwardsAIAct, many_hands_ai}. To illustrate, the ML development process will typically include some or all of the following (this is not a complete list):
\begin{itemize}
    \item \textbf{Training data} - this could be gathered and curated by multiple organisations and individual actors, such as anonymised health data \cite{training_db_cb} or publicly available datasets \cite{open_data_sets}
    \item \textbf{Software tools, libraries and services} - this includes off-the-shelf components produced in an open source environment, such as the Gazebo simulation environment \cite{gazebo} or PyTorch \cite{pytorch}, and cloud services for data storage and performing the training.
    \item \textbf{Dynamic system change} - the ML may continue to self-adapt after deployment \cite{EdwardsAIAct}
    \item \textbf{Informal requirements and incomplete verification criteria} - ML is often used for tasks that are difficult to formalise, e.g., "recognise road signs on UK roads in typical weather and light conditions", which make it difficult to train and verify across all scenarios it will encounter \cite{hawkins2021guidance,ashmore2021assuring}
    \item \textbf{Changing and dynamic deployment environment} - the AI-SCS can be deployed in a continuously changing environment, e.g., in the case of autonomous vehicles \cite{burton2020mind}.
\end{itemize}

Hence, during development and operation of an AI-SCS there will be multiple developers, engineers, suppliers of components and data, project managers, operators, investors, regulators etc. and it is infeasible to document and link all decisions they make which could contribute to safety. However, if we model key relationships and responsibilities with more clarity, we could understand and resolve some issues such as conflicts, duplicates, non-allocated responsibilities, and prevent safety problems in the future, as well as support accident investigation.

This "problem of many-hands" was introduced by Thompson in \cite{thompson_1980} and later studied for conventional high-criticality systems \cite{thompson_2017,nissembaum_account}. It can be summarised as the problem that, because many individuals and groups of individuals contribute to decisions, activities and outcomes in complex networks and organisations, it is difficult even in principle to determine who is responsible for them. Thompson uses the cases of the Banking crisis in 2007, and Gulf Oil spill in 2010 to illustrate the difficulty in finding accountable persons \cite{thompson_2017}. Additionally, he notes that some may unfairly avoid blame and others unfairly take blame. To avoid similar incidents, Thompson proposes the use of additional bodies with oversight responsibility and ensuring conflicting responsibilities are assigned to different entities.

The problem of many hands, and the difficulties in identifying responsible agents for machine learning in a non-safety environment is explored in \cite{many_hands_ai} by Cooper et al. and in \cite{cobbe_account} by Cobbe et al., noting the many different responsible agents. Cooper notes that the inevitability of ML bugs can be used to excuse responsibility for non-safety systems. We note that for SCS reducing bugs is standard practice to reduce risk, and should be evidenced as part of a SAC \cite{hawkins2013principles}. However, bugs will still remain despite being reduced systematically. Cooper also argues that responsibility should be designed into organisations while systems are developed. Using a role responsibility model, such as the one we propose, is one way this can be achieved for AI-SCS. 

\subsection{Legal pitfalls and due diligence}
\label{sec:soft_law}

One key aspect of responsibility is that of legal responsibility and liability following an accident. 

We previously noted engineers follow "good practice" for developing and operating traditional safety-critical systems, e.g., DO-178C for avionics \cite{178c}, ISO-14971 \cite{14971} for medical devices, or ISO-26262 \cite{26262} for electronic automotive vehicle components. Not following these could lead to sanction, even if the standards are not legally required, and hence is related to responsibility. One problem for AI-SCS is that there is limited consensus and little evidence of what constitutes good practice for developing safe ML \cite{ashmore2021assuring}, despite a large number of AI/ML safety standards \cite{AIStandards}, and huge volume of research in this area. Therefore, alternative means of demonstrating due diligence and good practice will be needed.

In the UK, the Health and Safety at Work etc. Act 1974 \cite{hse}and the subsequent Management of Health and Safety at Work Regulations of 1999 lay down the general legal obligations pertinent to ensuring the health and safety of both employees and people who may be affected by the activities of an organisation. The Act gives rise to criminal sanctions for organisational failures and emphasises the importance of proactive measures (i.e., risk assessments, health surveillance, and communication between employers and employees, amongst others) to ensure the overall health and safety of individuals. 
Based on the guidance issued by the Health and Safety Executive, proactive risk assessments play a vital role in safeguarding individuals by identifying potential harms and implementing measures to mitigate such risks from materialising. This practice is fundamental.

This emphasises the legal importance of due diligence practices. From a legal perspective, one way of describing the purpose of a SAC is a proactive means of justifying and explicitly detailing the roles and responsibilities of the various individuals involved in a system’s lifecycle. This clarifies their duties and outlines their subsequent responsibilities. The SAC works in two ways: both forward-looking and backward-looking. The intention is to assess potential risks from the outset and identify the relevant roles responsible for mitigating such harms. However, in the event something goes wrong, the SAC will be used in legal proceedings to assess how and where the risk materialised. In complex negligence-based cases, of specific pertinence to AI-SCS, industry standards and expert evidence may be used to assess the scope of duties, any deviations from the standard, and negligent conduct. Published material or guidance forming the general corpus of knowledge in the field may also be used as a reference and can be the deciding factor in liability claims. Importantly, if a person actively and diligently assesses risks, implements safety procedures, and ensures the relevant guidelines are followed, the person is said to have taken reasonable measures to ensure the safety of their conduct. In the event something undesirable occurs, it can be argued it would be unreasonable to find someone who diligently follows good safety practice guidelines (and the law, more generally) liable – so long as they have evidently carried out their responsibility, as reasonably practicable. 

\subsection{Clarifying types of responsibility for AI-SCS}
\label{sec:resp_precision}
The previous discussions have touched on a number of different notions of responsibility but has not defined them precisely, nor characterised their relationships. Nor has it identified a practical means to review and analyse responsibility relating to AI-SCS, particularly for stakeholders who may be developing a SAC.

The work of Porter et al. in \cite{unravelling_resp} draws upon Hart’s classic taxonomy of the different senses of responsibility \cite{hart} to identify and clarify different types of responsibility (role, causal, legal and moral). This breakdown of types can be used to describe different aspects of responsibility relating to AI-SCS development and operation. Role responsibility refers to the tasks, specific duties, and obligations that attach to particular roles. Causal responsibility is another way of referring to causation. If one is causally responsible for something, one is just a cause of it or a salient causal factor in its coming about. Legal liability responsibility includes being required to pay financial compensation, or be subject to a legal order, or face punishment \cite{porter}. Moral responsibility (at the heart of much of the debate about responsibility gaps e.g., \cite{Matthias_resp, tigard}) concerns whether an actor deserves blame or praise for an outcome. This can overlap with the other senses of responsibility, such as role responsibility. Causal responsibility is generally taken to be necessary but not sufficient for moral responsibility, but does not necessarily imply legal responsibility. One can be held vicariously responsible without causal contribution. Additionally, there is a distinction between moral accountability (where an actor is held to account) and moral attributability (where an actors contribution can be considered voluntary). This distinction is revisited in our examples (Sections \ref{sec:uber_case_study} and \ref{sec:diabetes_case_study}). 

In \cite{unravelling_resp}, Porter et al. present a formulation for representing responsibility as follows:
\begin{quote}
        \textit{Actor(A)\ is\ Responsible\ for\ Occurrence(O)}
\end{quote}

The \textit{Actor (A)} could be an AI-based system (such as a AI-SCS), an individual involved in development or operation, or an institution (e.g., company or organisation). \textit{Occurrences} are characterised as either \textit{Decisions}, \textit{Actions} or \textit{Omissions}. For each an asterisk (\textit{*}) is used to indicate where the \textit{Occurrence (O)} is attributed to an AI-based actor. For example, \textit{Automated Driving System (A) is (task)role-responsible for executing the dynamic driving task whilst activated*(O)}. We have used this formulation as a basis for modelling responsibility types.

Additionally, we need perform effective analysis of our model. In \cite{resp_model_sommer, resp_model_Lock} the authors present a responsibility analysis for high criticality systems. Their analysis is for operational roles, considering duties and tasks of human actors. They also provide a notation for the analysis which includes responsibilities, resources, agents, and different types of relationships. In \cite{resp_model_Lock} the authors use guidewords to prompt review of failures across the relationships, such as \textit{early, late, never}. In \cite{resp_model_sommer} (role) responsibility failure types including \textit{unassigned, duplicated, not communicated} are used. We have adapted their notation and analysis approach (see Section \ref{sec:method}), incorporating the types of responsibility described above, reinterpreting and adding to the guidewords and extending to include relationships. 

\subsection{Summary}
In summary, research in understanding responsibility for AI-SCS, has identified an number of responsibility related issues. Reviewing and documenting role responsibility during design of AI-SCS could prevent incidents in the future, e.g., when there are conflicting or missing responsibilities that lead to safety risks. Our contribution is to demonstrate a practical means of modelling and analysing responsibility for development and operation of an AI-SCS, to clarify responsibility, and, if possible, reduce responsibility issues. The analysis can be used as part of a SAC to demonstrate due diligence and also potentially as defence following an accident. Further, it could be used to investigate and prevent similar issues in the future, and avoid unfair blame. 
\section{Method}
\label{sec:method}
In this section we describe our responsibility modelling notation, and give an overview of how to develop and analyse models. Sections \ref{sec:uber_case_study} and \ref{sec:diabetes_case_study} show two different examples of it being used.
\subsection{Notation}
\begin{figure}[htbp]
\centering
 \includegraphics[scale=0.3]{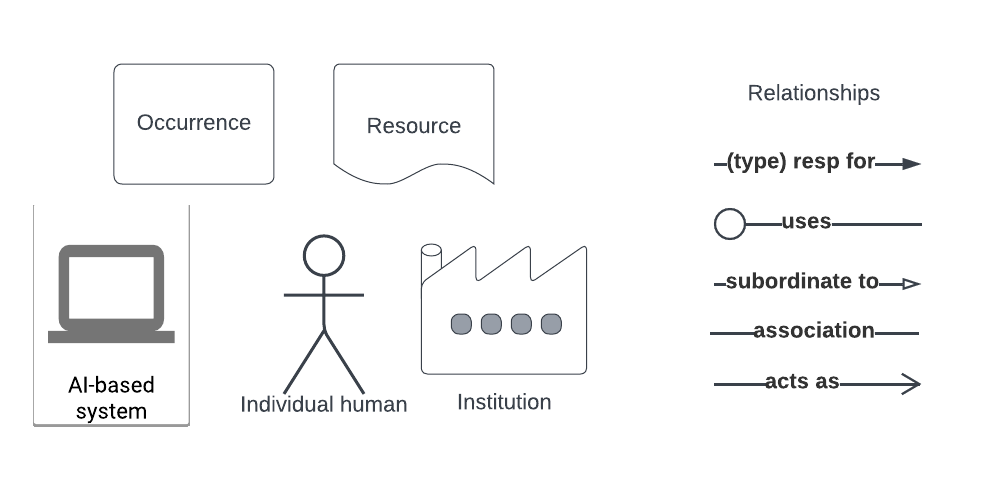}
  \caption{Responsibility model elements}
  \label{fig:resp_model_elements}
\end{figure}

Our notation is an adapted version of the task responsibility notation presented by Lock in \cite{resp_model_Lock}, incorporating the responsibility concepts provided by Porter et al. in \cite{unravelling_resp}. This combination ensures that the relationships between responsibilities can be modelled, whilst adding the specificity and nuance of the responsibility concepts from \cite{unravelling_resp}. The model elements are shown in Figure \ref{fig:resp_model_elements}. The three different actor types identified in \cite{unravelling_resp} are presented (AI-enabled or AI-based system, individuals and institutions). What is referred to as \textit{Occurrence} by Porter et al., and \textit{Responsibility} by Lock is represented by a rectangle. Resources are represented using the standard flow chart symbol for documents. We include resources to allow us to model specific outputs from an \textit{Occurrence}. Actors may use a resource immediately (for example, a prediction) or later  (for example, an Off the Shelf software library used during development).

The relationships between the elements are as follows:

\begin{itemize}
    \item \textbf{(type) responsibility for} - this specifies variations on the four types of responsibility described in section \ref{sec:resp_precision} 
    \item \textbf{uses} - we have added this relationship to indicate where a resource or occurrence is used by another actor. This can add insight where problems with the resource (which may be due to prior issues undertaking the task to manage it) may have safety implications. We note that although a resource may be \textit{used} it may not necessarily be \textit{required} for an actor to complete their task.
    \item \textbf{subordinate to} - this represents where there is a power relationship or hierarchy between elements
    \item \textbf{association} - this is used where there is a non-specific relationship between elements, for example, a resource associated with an occurrence
    \item \textbf{acts as} - where one actor or element acts in place of another
\end{itemize}

\begin{figure}[htbp]
\centering
 \includegraphics[scale=0.25]{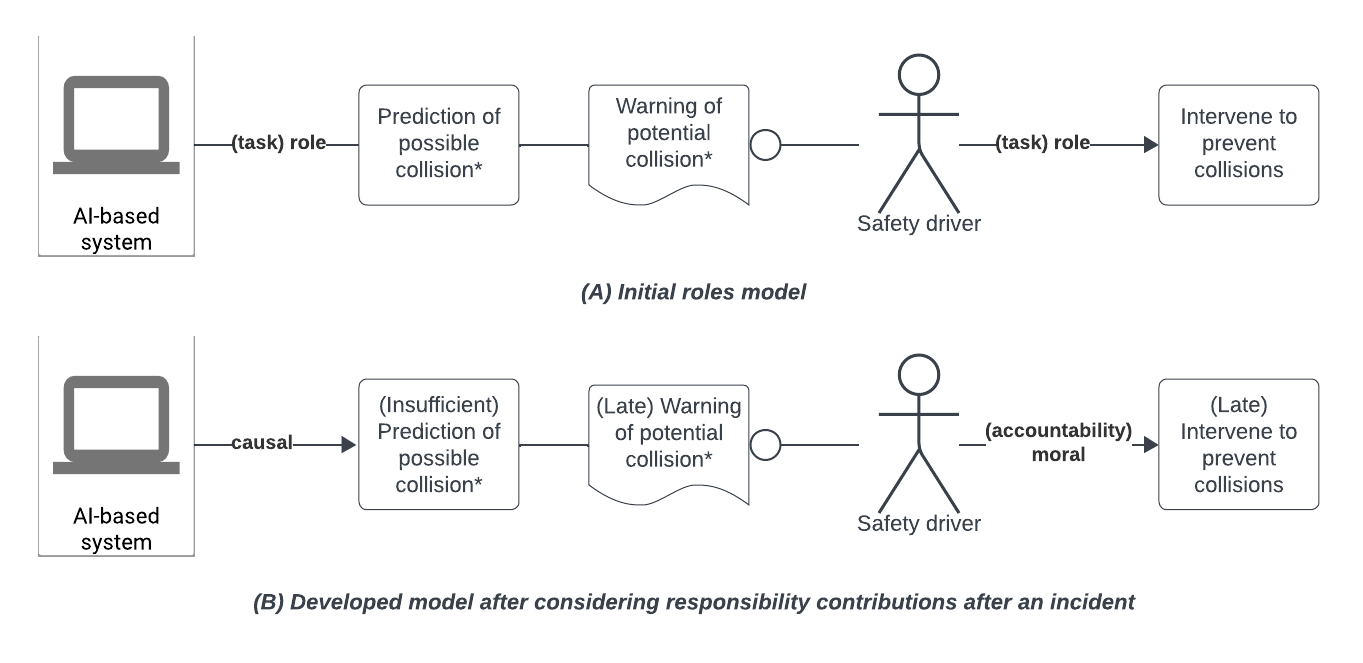}
  \caption{Development of a simple responsibility model}
  \label{fig:simple_resp_eg}
\end{figure}

We illustrate the concepts in Figure \ref{fig:simple_resp_eg}, where an AI-based system has the \textit{(task)role} of monitoring whether there is a pedestrian on a potential collision path. If so it creates a collision warning (resource) which is used by the Safety Driver. The safety driver is responsible for intervening to try and prevent a collision when a warning is issued (if necessary). 

The initial model is shown in the top line (A) where the actors and their tasks are described. Considering failures allows us to see the impact on actors, and how they may perform their role. For example, if the warning is late due to poor performance of the AI-based system in its monitoring task, this may mean a late intervention. In the bottom row (B), we show the responsibility contributions following a collision. The AI-based system has a causal link to the insufficient prediction and associated warning, whereas the safety driver may be blamed for late intervention. The AI-based system is not a legal person and is not considered to be a moral agent, hence cannot be held liable or morally responsible \cite{morgan_book, porter}. However, the safety driver may be found morally accountable or even liable following the crash, even if the AI-based system causally contributed to their actions. This highlights a responsibility gap between a causal factor and the driver.

We can extend this model and analysis to include the actor(s) responsible for designing the AI-SCS, or who provided training data, etc., and build a chain of actors and relationships. This can help prevent issues of AI technology shortfalls by identifying tasks to reduce risk, as well as clarifying roles of which manage and contribute to risk. Further, the model can be used to help demonstrate there was no responsibility shortfall within a SAC.

In a forwards-looking analysis there may be many different responsibility models, with different hypothetical outcomes, dependent on the issues uncovered. For example, the \textit{Warning of potential collision*} in Figure \ref{fig:simple_resp_eg} could be \textit{Late?} or \textit{Missing?}, with different consequences.

\subsection{Creating and analysing the model}

There are two main phases for creating and analysing the model. First, information on the elements needed to create the diagram is gathered, concentrating on identifying relevant actors and roles. Second, an iterative analysis of the model is performed either during development or after an incident. These phases should be performed by teams of experts with appropriate specialist knowledge to the AI-SCS (e.g., clinicians, ML software engineers, safety experts, legal experts).

\subsubsection{Gathering information}
The responsibility model is built with the following steps:

    \begin{itemize}
        \item Identify key elements:
        \begin{itemize}
            \item List key operational and development actors (human, institutions and AI-based)
            \item List key resources and occurrences used during operation: such as system status outputs, information, interacting tasks, compliance and\footnote{Compliance duties relate to the use of safety standards and established "good practice". Whilst not legally mandated, lack of compliance could lead to sanction (see Section \ref{sec:soft_law}).} legal duties
            \item Using the previous data, list additional development roles, resources and occurrences which relate to them
        \end{itemize}
        \item Characterise the responsibility relationships between the elements
        \item Arrange elements into graphical model
    \end{itemize}

A few comments on this process. If this is a forwards-looking model, we assume that the main responsibility types will be moral obligations and duties (for example, codes of conduct), legal duties (e.g., duty holders or legislative roles) and task roles (e.g., typical development activities). Backwards-looking models will include different responsibility types such as causal, moral attributions and liability, but we will speculate on these during analysis. As noted, review and analysis will support the model's development and iteration.

Each actor may embody different types of responsibility (leading to occurrences and resources), and each occurrence and resource could potentially have multiple related tasks. For example, a single resource could be managed by more than one actor, and in different ways, such as a car being built by a manufacturer and maintained by a dealer.

\subsubsection{Reviewing and Analysing the model}
The model will require review and analysis to determine potential issues, such as risk reducing tasks with no assigned actors or actors who have overloaded roles and a high burden of risk. 

The initial aim of the analysis is to build confidence that all risk related roles have been identified and allocated, as shown in section \ref{sec:diabetes_case_study}. For a backwards-looking analysis, the model can help determine where role responsibility issues led to an incident or accident. An example is shown in section \ref{sec:uber_case_study}.

Some established safety analyses, such as Hazard and Operability Studies (HAZOPS) \cite{resp_model_Lock, pumfrey, DCP_safecomp} use guidewords to suggest failure conditions in SCS, e.g., "Lack of" meaning a communications message is not sent, or "too much" meaning the addition of too many chemicals to an industrial process. As these guidewords are proven, we use these when considering technical issues with \textit{resources}.

However, no established method exists for our responsibility analysis. Again, taking inspiration from the work of Lock \cite{resp_model_Lock} and Baxter \cite{resp_model_sommer} we have developed a set of guidewords used to prompt discussion and identification of issues for occurrences. These are now described, with examples of how they might prompt discussion.

\begin{itemize}
    \item \textbf{Insufficient/Partial} - where an Occurrence has happened but it doesn't meet the requirements in full
    \item \textbf{Misassigned} - where it is not appropriate to assign a particular Occurrence to a particular actor
    \item \textbf{Overloaded} - where an actor may have too many tasks, or they are too burdensome
    \item \textbf{Duplicated/Conflict} - this indicates two or more actors performing the same or similar occurrence, noting that they may conflict
    \item \textbf{Missing/Never/Omission} - where an occurrence is not provided
    \item \textbf{Early/late/ordering} - where an occurrence is provided but there is an issue with the sequencing which causes a problem
    \item \textbf{Incorrect/Value} - where there is some other issue with the occurrence, such as it being performed incorrectly
\end{itemize}

For each of the occurrences and related resources, the safety impact of an issue needs to be assessed. For example, an actor may not be appropriately trained to perform a task (\textit{misassigned}), meaning their outputs would not be sufficient, or it may not be assigned (\textit{missing}). Another scenario is that we cannot provide guarantees in the quality of a resource (\textit{insufficient} e.g., where there are many actors i.e., many hands, producing open source software). The analysis may uncover additional tasks which either alter existing occurrences/resources or create new ones. For example, an additional assurance task needed to assure software should be to be assigned to an appropriate actor, or a duplicated task may need clarification of who should perform it.

\section{Example 1 - Fatal collision with an Autonomous Vehicle in Tempe, Arizona}
\label{sec:uber_case_study}
In this section we present a backwards-looking analysis to illustrate the modelling concepts and methodology in more detail. The example is the fatal collision of an Uber Advanced Technologies Group (ATG) vehicle with an automated driving system (ADS) and a pedestrian pushing a bicycle across a highway in Tempe, Arizona in 2018. The National Transportation Safety Board (NTSB) issued an accident report \cite{uber} describing contributory factors, and highlighting a number of areas where there were safety shortfalls. Uber ATG have since published a report describing improved safety processes to address some of the shortfalls identified \cite{uber_internal}. A detailed sociotechnical analysis of the accident can be found in \cite{macrae_analysis} which points at many undesirable organisational factors such as undue pressure to deliver and update a working solution, learning lag when reviewing in service data due to too much information, push for technical capability constraints to prioritise normal performance over safety, and over-reliance on the safety driver's reliability over long periods of time monitoring for hazards. These reports were the main sources of information for our analysis.

Uber ATG were responsible for developing the ADS, which was an adaptation added to a Sport Utility Vehicle (SUV) from a third-party supplier (Volvo). The ADS used a number of ML components to detect and classify objects in the vehicle's path, and also to predict those objects trajectories. The findings of the accident report noted that the classifier failed to consistently categorise the object as a pedestrian with a bicycle. Each time the classification changed, information on the pedestrian's path was lost, and so the ADS could not predict the pedestrian's path correctly to provide a collision warning. The safety driver in the vehicle was trained to disengage the ADS in emergency situations and take avoidance action. She was found to be distracted during the time of the accident. It was noted that there was inadequate regulatory control over operation of an AV.

In 2019, prosecutors said Uber ATG was not criminally liable in the crash. Uber ATG also settled the civil case brought by the victim’s family out of court. Uber ATG preferred to use the word “resolved”, in place of settled, as they did not admit liability \cite{Fulbrook_uber}, although settlements which compromise a dispute often contain provisions that a party does not accept liability. In our opinion, the out of court settlement represents a missed opportunity to test important questions of civil liability for incidents involving AI-SCS.

\subsection{Producing and analysing the collision responsibility model}

\begin{table*}
  %\hyphenpenalty 10000
  %\renewcommand{\arraystretch}{1.2}
  \caption{Example Uber ATG Tempe accident responsibilities following legal proceedings conclusion \cite{uber, TAS_symp}}
  \label{tab:uber}
  \small
  \begin{tabular}{p{2.5cm}p{1.75cm}p{2.5cm}p{1.5cm}p{1.5cm}p{2cm}}
    %\toprule
    \hline
    Finding & Related Roles & Compliance \ duties & Legal\ Duties & Causal \ Contrb & Liability\\
    %\midrule
    \hline
        Lack of driver attentiveness & Safety driver & Compliance to training & Yes & Yes & Pleaded guilty to charge of "endangerment" \cite{UberOverSeattle, UberOverWired}  \\ %\hline
        Limited monitoring of driver attentiveness & Uber ATG & None identified & None & Yes & Not tested  \\ %\hline 
        Emergency response & Emergency services & N/A & Yes & None & None  \\ %\hline
        Pedestrian behaviour & Third party & N/A & Illegally crossed road & Yes & None  \\ %\hline
        Limited risk analysis of experimental systems & Uber ATG & Established good practice for risk assessment & None known & Yes & Not tested.  \\ %\hline
        Deactivated automated braking system & Uber ATG & Established good practice for risk assessment & None known & Yes & Not tested  \\ %\hline
        Lack of safety culture & Uber ATG & Established good safety practice & None known & Yes & Not tested  \\ %\hline
        Lack of state/ federal oversight for AVs & Regulator/ lawmakers & Not available & Not available & Yes & None  \\ 
        
  %\bottomrule
\end{tabular}
\end{table*}

\begin{figure}[htbp]
\centering
  \includegraphics[width=13.5cm]{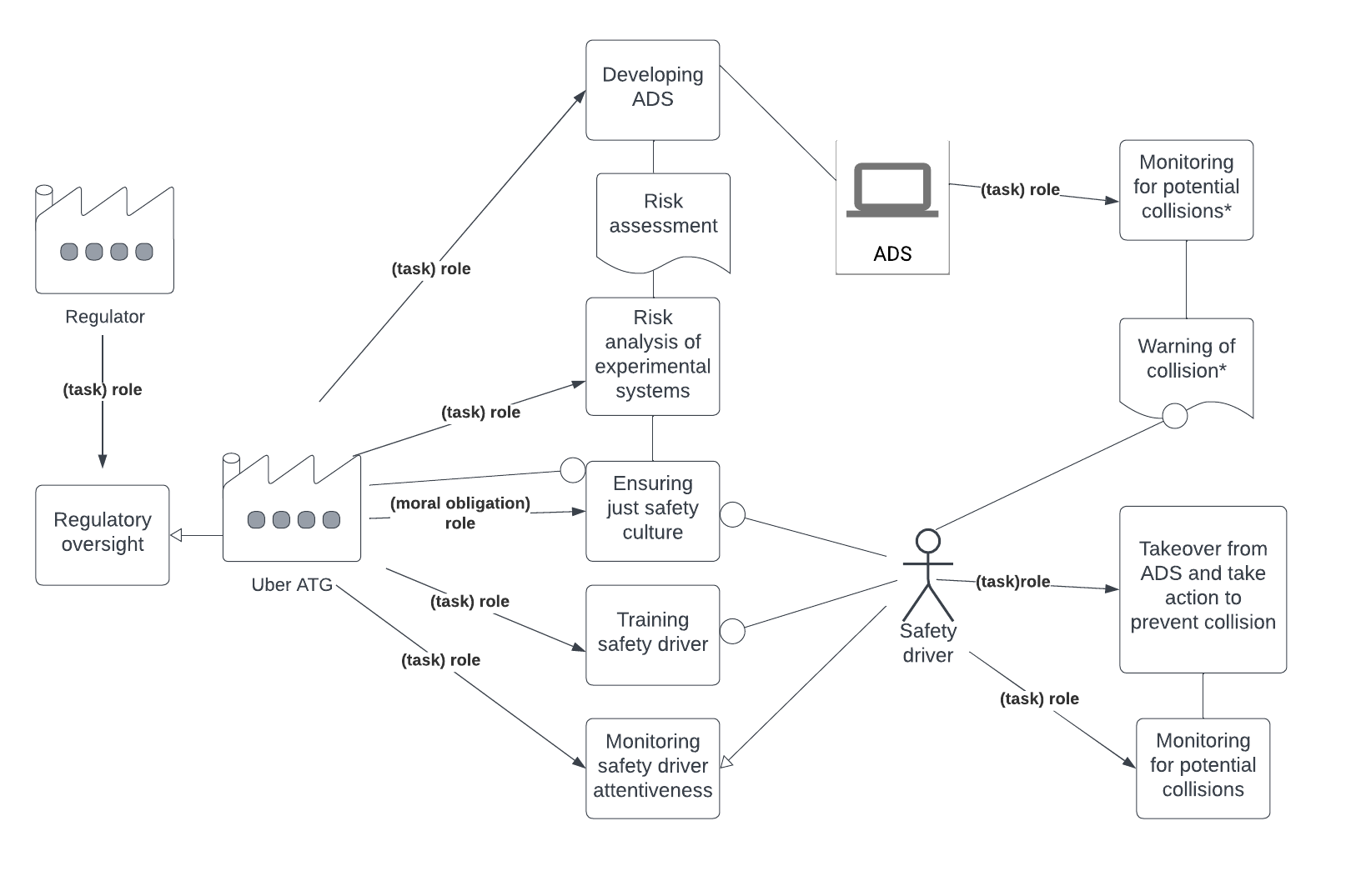}
  \caption{Initial responsibility model showing actors and roles from Uber ATG crash in Tempe}
  \label{fig:uber_resp_tasking}
\end{figure}

\begin{figure}[htbp]
\centering
 \includegraphics[width=13.5cm]{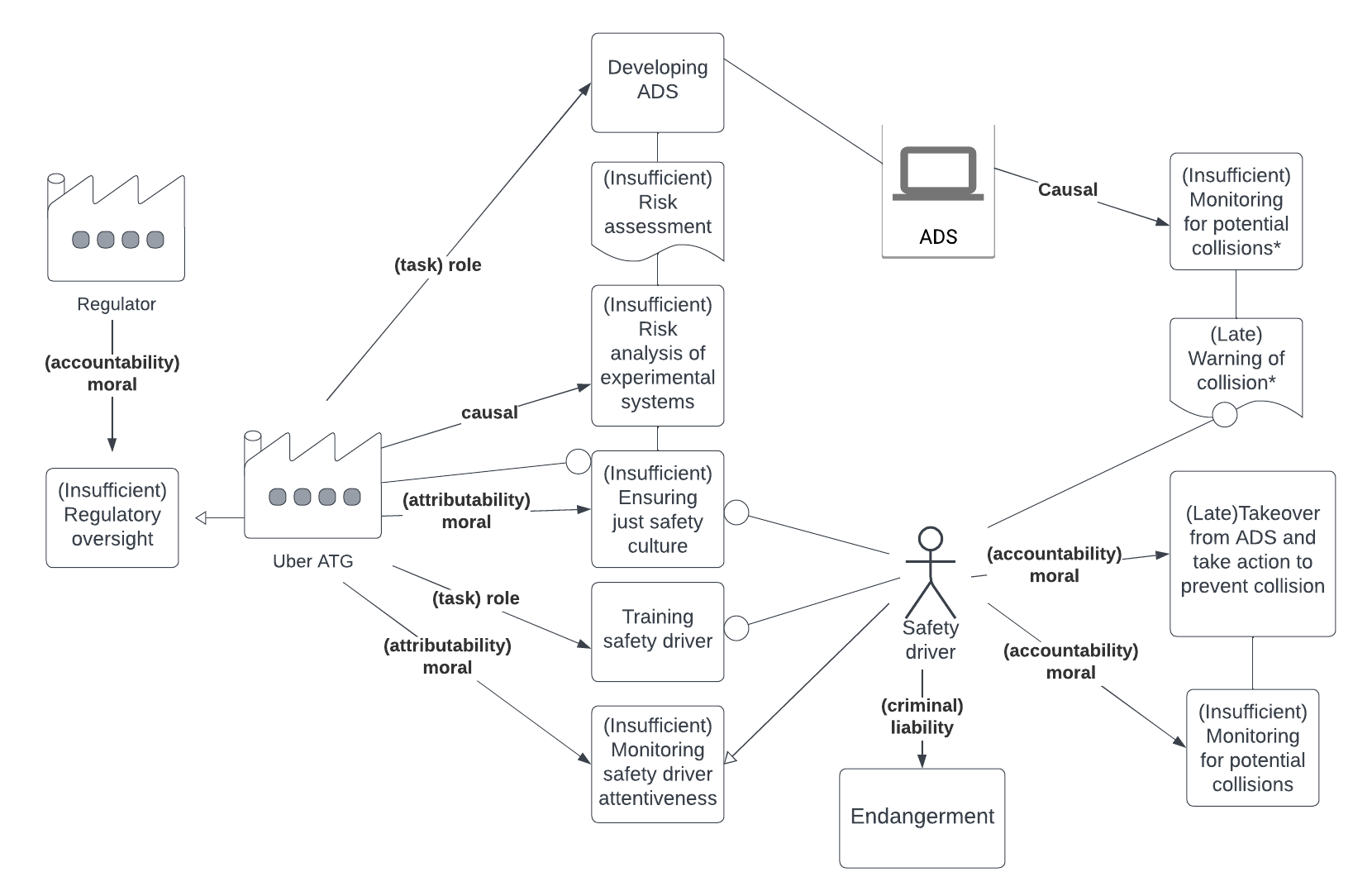}
  \caption{Extended Uber Tempe responsibility model highlighting some responsibility findings from Uber ATG crash in Tempe \cite{uber}}
  \label{fig:uber_resp_findings}
\end{figure}

In previous work \cite{TAS_symp} we presented an analysis of the Uber Tempe accident looking at each responsibility type and related actor individually. Our work summarised the different safety controls and mitigations which were in place at the time of the accident, their effectiveness, and who was assigned the role of implementing them either for development or operation. Since publishing the legal case against the safety driver has been completed, as she pleaded guilty to "endangerment" \cite{UberOverSeattle,UberOverWired, Shepardson_uber}. Here we include an updated analysis with the legal outcome, introducing our modelling notation and showing formal relationships between actors. In Table \ref{tab:uber} we list some of the report findings, relevant roles, their compliance and legal duties, whether these duties could be a causal factor in the accident and their ultimate liability. We present two models of role responsibility, illustrating some relationships to visualise the relationships between the different roles which led to the accident. The first model shows the actors, and linked role responsibility tasks prior to the incident (Figure \ref{fig:uber_resp_tasking}), and the second model shows the findings after the accident (Figure \ref{fig:uber_resp_findings}).

The models were developed as per the method outlined in section \ref{sec:method}, starting with an initial model that includes the different actors, and their roles and relationships, prior to an analysis. We have modelled their relationships based on our review of accident findings. The initial model shows a subset of the actors with the tasks assigned to them in Figure \ref{fig:uber_resp_tasking}. 

A few comments on the initial model (Figure \ref{fig:uber_resp_tasking}). The production of a safety assessment was voluntary in accordance with the regulatory framework, but Uber ATG were \textit{subordinate to} the regulator and so we have presented their relationship as such. Compliance and legal obligations should be considered for every AI-SCS responsibility model due to their importance in development and operation, and specifically due to grey areas around legal liability of AI based systems (see section \ref{sec:soft_law}). 

We have modelled \textit{Ensuring just safety culture} as a \textit{(moral obligation) role}. This reflects the obligation to embed safety concerns within the organisation, in accordance with general societal expectations and norms. No individual actor is named as responsible, as it is a collective obligation, although we note that individual degree of responsibility may depend on authority and seniority \cite{dekker}\cite{safety_culture_trans}. Other actors have \textit{(task) roles} related to development (e.g., \textit{risk assessment}) and operational occurrences and resources (e.g, \textit{Monitoring safety driver attentiveness} and \textit{Warning of collision} from the ADS). They are not named individually in public materials, so Uber ATG is presented as an institution. The ADS and safety drivers are presented as separate actors as they have distinct functional role during operation.

We have not included the pedestrian as an actor in this model. We could include third parties impacted by the \textit{occurrences}, for example, near misses or collisions with other road users, loss of jobs for drivers due to autonomous taxis. As these actors have no task in this situation, we have not included them. Arguably, other road users do bear some responsibility though, e.g., to follow the rules of the road. A future consideration is how to effectively incorporate impacted third parties where their \textit{(task)role} may not directly interconnect.

Our model has been adapted in Figure \ref{fig:uber_resp_findings} to highlight some of the findings of the accident report and other publicly available material. First, a note of a guideword encapsulating a shortfall is included in each of the \textit{occurrences}, e.g., \textit{Warning of collision} becomes \textit{(Late) Warning of collision}. Second, where appropriate we have updated the relationship arrow annotations to reflect the findings, e.g., \textit{(task) role} becoming \textit{causal}.

Again, we have a few comments on this model. In order for someone to be held either morally accountable or attributable to an \textit{occurrence}, it is considered a necessary condition for them to both have moral agency and some causal contribution \cite{unravelling_resp,hart}. The ADS had a causal contribution, via \textit{(Late) Warning of collision} but has no moral agency. 

In accordance with Porter et al. in \cite{unravelling_resp}, we can consider "moral responsibility as attributability ... when [actor] has voluntarily performed or voluntarily caused [occurrence]", and wasn't under extreme pressure or duress. With this in mind we consider that there is an \textit{(attributability) moral} responsibility relation with the \textit{(Insufficient) Ensuring just safety culture} for Uber ATG. However, we have annotated a \textit{causal} responsibility link to \textit{(Insufficient) risk analysis of experimental systems}. McCrae \cite{macrae_analysis} notes that there was both pressure to deliver (based on fears that the organisation's existence was at stake) and staff were disempowered or discouraged from speaking about safety concerns. Hence, we consider the insufficient risk analysis was possibly not voluntary, but was arguably due to significant pressure. As noted, Uber ATG are represented as an institution, with no specified individual actors as this information is not available. This is an example of the many hands problem (section \ref{sec:many_hands}) where we don't have traceability to individual decisions or actions. 

We also consider accountability, i.e., where someone faces the reactions and responses of the wider community for their actions. This includes blame or praise, for example, the emergency services were praised as providing "appropriate and efficient emergency medical response" \cite{uber}. From the finding of the legal case \cite{UberOverSeattle,UberOverWired, Shepardson_uber} and the accident report, we show that the safety driver's role can be considered blameworthy as \textit{(accountability) moral} for \textit{(Insufficient) Monitoring for potential collisions} and the late intervention. We have added the \textit{(criminal) liability} occurrence of \textit{Endangerment} to our model to reflect the legal case.

The addition of criteria to our analysis method for determining issues such as accountability and attributability is an avenue for further research, as it may be highly subjective and based on relative importance of causal contributions.

Uber ATG were found not to have monitored the safety driver's awareness, due to underestimation of automation complacency, i.e., where the safety driver trusts an automated system (the ADS) to an extent that they pay less attention. Also, there was a complex chain of related occurrences from problems with risk assessment, impacting on the ADS performance, leading to a late warning from the ADS to the driver. This removed a layer of design mitigation that should have reduced the risk of operating the vehicle, and reduced the cognitive load for the safety driver. A more prompt warning would provide mitigation against automation complacency, given more time for the safety driver to intervene, and also allowed for autonomous action to avoid collision. Arguably, this demonstrates where the burden of risk has increased on the safety driver due to shortfalls in how other actors performed in their roles.

The second model (Figure \ref{fig:uber_resp_findings}) illustrates many nuances of responsibility, and how the conduct by which some \textit{(task) roles} were performed led to an accident via relationships with other tasks. We note though, that our analysis is based on pre-existing knowledge of the accident. In the next section we explore whether the method can be used identify potential responsibility problems without prior knowledge, and support a SAC justifying they have been adequately managed.
\section{Example 2 - Diabetes co-morbidity clinical decision support AI-SCS}
\label{sec:diabetes_case_study}

Our second example is an AI-based diabetes co-morbidity predictor (DCP) the authors of this paper are developing for treatment of patients with Type II diabetes \cite{ozturk, DCP_safecomp}. The AI predicts a patient’s risk of developing a diabetes co-morbidity or having a potentially catastrophic event, such as a heart attack, within the next six months. It is a decision-support system used during a patient consultation by the clinician, and has been built using an ensemble of ML components \cite{ozturk} to process recent patient test data and make a prediction of "High" or "Low" risk of a particular condition. The clinician can choose not to act on predictions made by DCP if they disagree with its assessment or if they have other information, for example, about the patient’s preferred treatment, which make alternative courses of action more appropriate. The DCP is intended to function as an independent ‘second opinion’ during consultations. Previously, we identified that there are many complex issues arising from the public data sets used for training \cite{training_db_cb}, which could lead to misleading outputs from the system \cite{DCP_safecomp}. These can be partially mitigated by careful curation of the training data by the AI developer, however the impact of some issues (e.g., bias due to lack of data from certain patient demographics) cannot be easily reduced, nor can we exhaustively test the AI as there isn't a formal specification for required behaviour.

\begin{figure}
    \centering
    \includegraphics[width=0.8\linewidth]{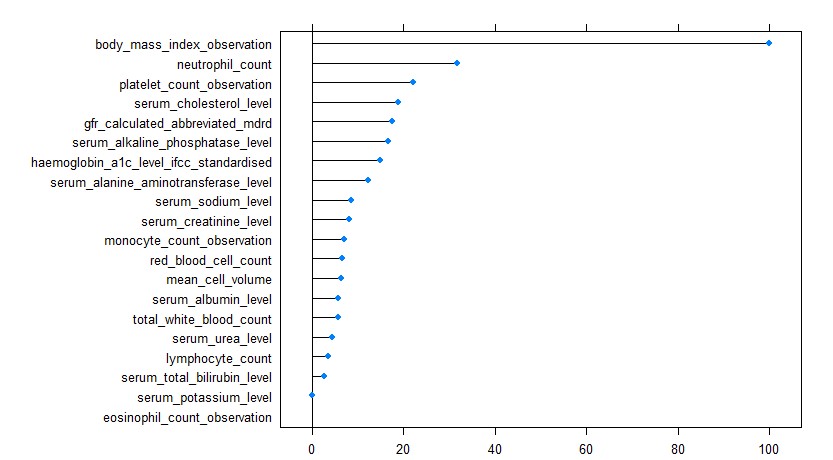}
    \caption{DCP explainability example showing influence of FOIs \cite{DCP_safecomp}}
    \label{fig:foi}
\end{figure}

The DCP additionally provides explanation data. This lists the patient data features of importance (FOIs) for a particular prediction with values indicating their influence on the output recommendation. An example is shown in Figure \ref{fig:foi}, where Body Mass Index (BMI) is shown as the most influential factor for hypertension diagnosis.

\subsection{Producing and analysing the DCP responsibility model}
\subsubsection{Producing the initial responsibility model}

This section discusses development of the initial model (Figure \ref{fig:dcp_rr}). As discussed in section \ref{sec:background}, we include the AI development lifecycle in our model to understand the impact of development issues on operational risk. This means we have included actors influencing the ML development process, regulators, and operating roles. Then we have considered each of their outputs as resources and/or occurrences.

The initial model is in Figure \ref{fig:dcp_rr}. A number of the human actors (\textit{Clinical Staff}) have the qualifier \textit{1..N}. This indicates where there are many actors performing this role, which we do not have detailed information for, and they are not attached to a specific institution. This is another example of the many hands issue discussed in section \ref{sec:many_hands}. For example, the \textit{Training patient database} is a resource that has been contributed to over many years, by different clinics with different unknown staff, each performing a \textit{Maintaining patient database} occurrence. The \textit{AI software tools} have also been developed by many different actors; in this case we assume they have come from specific software development institutions, for example, commercial tools. We could extend the model to consider each of these individually if valuable. 

For regulatory oversight we have listed the health regulator (e.g., Medicines and Healthcare products Regulatory Agency (MHRA) in the UK) and National Institute for Health and Care Excellence (NICE). MHRA provide approval for medical device safety assessments, and NICE provide clinical guidelines which we used to assist in determining the high/low risk of different co-morbidities. We have a node for \textit{AI development good practice,} which does not currently have an associated actor with the task of development, due to the lack of established AI specific medical device guidance. There are, for example, guidelines being developed by the U.S. Food and Drug Administration (FDA), but these are in draft and not proven \cite{FDA_AI_plan}. 

The operational actors (right hand side of the diagram), include \textit{Clinical Staff} responsible for maintaining non-electronic and electronic records for the patient. Additionally, we include the clinician performing a consultation with a patient, and the DCP. Unlike the previous example in section \ref{sec:uber_case_study}, the impacted party is included, i.e., the type II diabetes patient, as they have a specific role during consultation. Here, the clinician will interact with the patient, asking them questions about their condition and general health. We also include that the clinician has a \textit{(moral obligation) role} for \textit{Duty of care}.

\subsubsection{Analysing and developing the model}
\begin{figure}[htbp]
\centering
  \includegraphics[scale=0.24, angle =270]{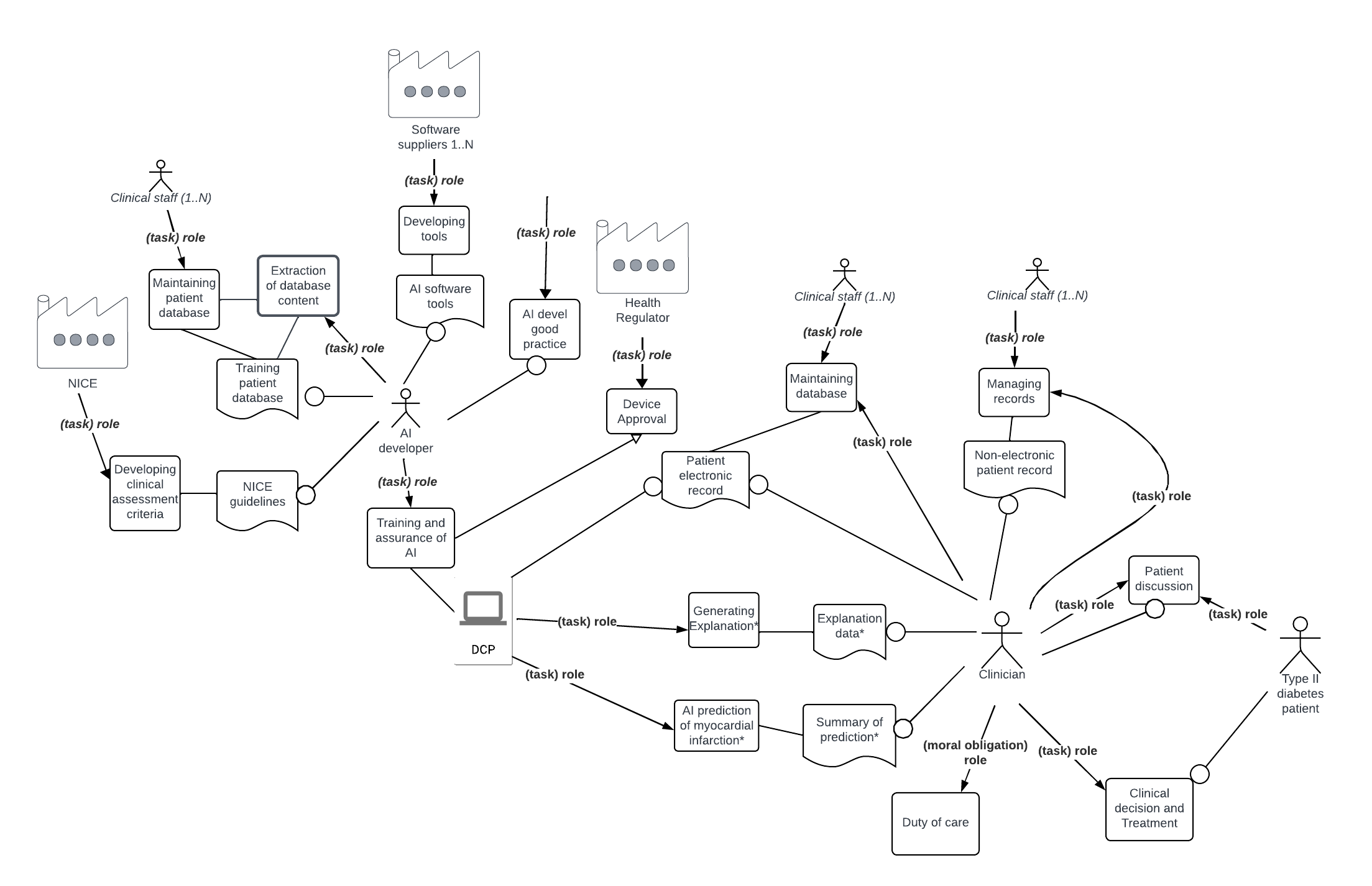}
  \caption{DCP (task)role relationships}
  \Description{Illustration of DCP initial Role resp model}
  \label{fig:dcp_rr}
\end{figure}

\begin{figure}[htbp]
\centering
  \includegraphics[scale=0.24, angle =270]{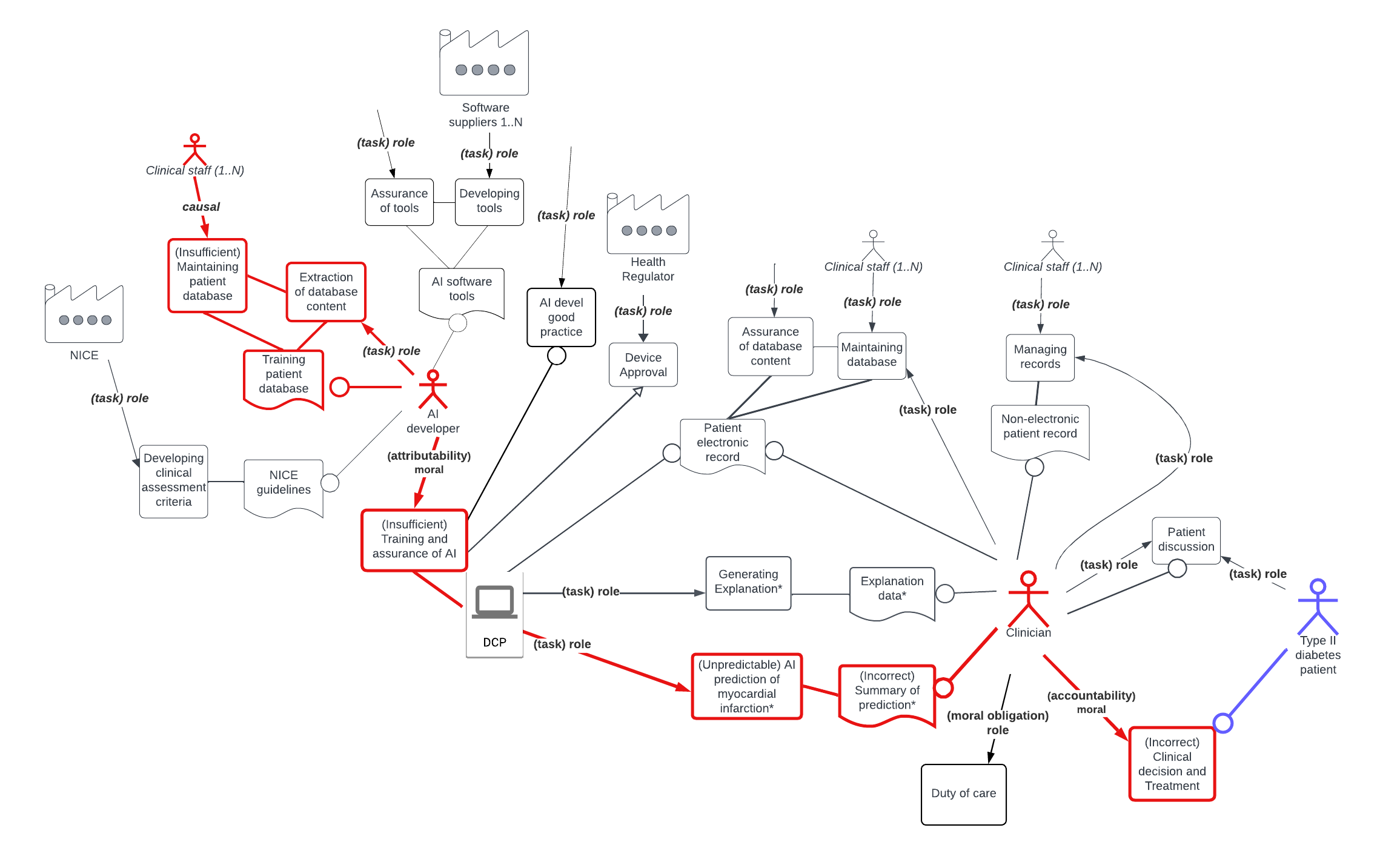}
  \caption{Revised DCP model (red highlighted path showing operational safety impact of training data errors)}
  \Description{Illustration of DCP iterated Role resp model}
  \label{fig:dcp_rr_adapt}
\end{figure}

In this section we demonstrate how a forwards-looking analysis of Figure \ref{fig:dcp_rr} helps develop the revised model (Figure \ref{fig:dcp_rr_adapt}). As described in section \ref{sec:method}, we need to consider hypothetical issues with both occurrences and resources, and think about how they may impact on safety of the AI-SCS. This should consider safety impact on the performance of the AI-SCS and on interacting actors. Issues may include, missing occurrences, actors with conflicting or overloaded roles, insufficient or late resource provision and so on. An extract of the HAZOP style analysis is shown in Table \ref{tab:dcp_hazop}. 

When applying the guidewords we have proposed some additional tasks (i.e., occurrences) to mitigate or prevent potential safety issues. For example, in the highlighted path in Figure \ref{fig:dcp_rr_adapt}, issues with the training data  have \textit{causal} impact on performance of the DCP. This is shown as eventually (potentially) contributing to \textit{(Incorrect) Clinical Decision and Treatment}. Mitigation measures can be assigned to the AI Developer, by improving how data is processed and extracted prior to training. In Figure \ref{fig:dcp_rr_adapt}, we hypothesise that by not performing task they are (\textit{(attributability) moral}) responsible for \textit{(Insufficient) Training and Assurance of AI}. However, the clinician would likely be held accountable (\textit{(accountability) moral}) for the incorrect diagnosis. The Type II diabetes patient would be affected. As previously discussed, stricter criteria to determine issues like moral accountability and attributability are an avenue for further research.

We need to assign the task of assuring the AI software tools. The chain of impact of bugs in tools is similar to that for training data quality issues. Bugs could remain undetected throughout development and into operation, affecting the predictions and explanations from the DCP, and thus impacting the decision making of the clinician. There are several reasons for this. First, the software tools may be complex and not designed for safety-critical use (for example, generic library tools used to manage and alter data, perform training of multiple neural network models etc.), making it difficult to assure effectively. Second, there may be no access to source code or documentation. Third, we cannot perform complete testing of the DCP to find bugs, due to the enormous variation of potential patient data and uncertain requirements for accurate predictions. A detailed causal impact analysis, considering impact, size of contribution and likelihood, for these chains forms an avenue for further research. As previously noted, several models may be needed for different outcomes.

Another potential issue identified by applying the guidewords, is whether explainability information provides useful insight into the AI prediction for the clinician \cite{jia2023need} \cite{jia2022role}. As discussed, a visualisation of the relative influence of FOIs on the prediction can be provided (see Figure \ref{fig:foi}). However, some FOIs could influence the DCP prediction in similar proportions whether it indicates high or low risk of a heart attack, or be close to NICE guideline boundaries. The clinician may choose not to use explainability data if it is ambiguous.

Table \ref{tab:dcp_hazop} proposes that incorrect (i.e., false negative/false positive) and conflicting predictions from the DCP are mitigated by the Clinician, by discussion with the Patient, other information, and use of their clinical expertise. 
This has added additional tasks, potentially overloading them and extending consultation times. Additionally, as the clinician is mitigating many potential problems with the AI partially introduced by other actors, we could argue their burden of risk has increased, rather than being reduced, by the use of the DCP. Again, we are concerned about them becoming a "liability sink" in the future. The model and analysis provides detail and clarity on where additional risks have come from.

\begin{table*}
    \centering
    \label{tab:dcp_review}
    \small
    \begin{tabular}{p{1.2cm}p{1.35cm}p{1.35cm}p{1.35cm}p{1.1cm}p{2.6cm} p{2.5cm}}
    \hline
       Occ.  & Resource(s)  & \textit{(task) role} Actor & \textit{Uses} Actor  & G/word & Issue & Mitigation \\
    \hline
      Develops tools & SW tools  & Software supps (1..N)  & DCP \newline developer  & Insuff. & Tool quality is unknown. Bugs impact on DCP performance. & Perform assurance assessment task to assess potential issues \\
      Maintain database & Training dataset  & Clinical staff (1..N)  & AI \newline developer  & Insuff. & Data poorly \newline distributed, missing values. DCP outputs are insufficient, e.g., perform poorly on patients matching missing elements. & Perform training data quality assessment and compensate where possible \\
      Training and assurance of AI & Prediction from DCP  & AI Dev.  & Clinician  &  Insufficient & Inconsistent and misleading performance of DCP  & Use explainability, follow up patient progress to assess DCP performance over time \\
      Training and assurance of AI & Prediction from DCP & AI Dev.  & Clinician &  Conflict & DCP provides FP/FN, impacts clinical decisions & Use explanability, follow up patient progress to assess DCP performance over time, prioritise clinical expertise \\
      Training and assurance of AI & Prediction from DCP & AI Dev.  & Clinician &  Conflict & DCP provides TP/TN but conflicts with clinician & Use explanability, follow up patient progress to assess DCP performance over time, prioritise clinical expertise\\
      Clinical decision and treatment  & N/A & Clinician & Type II diabetes patient & Incorrect & Wrong treatment recommendation due to influence of DCP prediction (FN/FP) & Patient discussion, use explainability, prioritise clinical expertise \\
      Generating explanation  & Explaination data & DCP & Clinician & Insuff. & The FOI data adds limited insight into the prediction & Patient discussion, prioritise clinical expertise \\
    \hline
    \end{tabular}
    \caption{Extract of the model HAZOP analysis for the DCP}
    \label{tab:dcp_hazop}
\end{table*}

\section{Discussion and future work}
\label{sec:conc}

There are many concerns around responsibility relating to the development and operation of AI-SCS, including "responsibility gaps", where it is difficult to ascribe blame or determine liability for unsafe behaviour of an AI agent, "the problem of many hands" where it is hard to identify responsible individuals, and "liability sinks" where operators have increased burden of risk when using an AI-SCS. We argued that modelling and analysing responsibilities during design of AI-SCS provides the potential to reduce risk, e.g., by ensuring there are no conflicting or missing responsibilities. We also hope to avoid unfair or disproportionate blame being placed on operators or developers. 

We presented a practical method for responsibility modelling which can support a SAC, required for most AI-SCS. We presented two different examples, demonstrating how modelling responsibility provides clarity on different roles and responsibilities in development of a AI-SCS, as well as the impact on operational safety, and the potential risk burden placed upon operators.

We first demonstrated how to develop a responsibility model, using prior knowledge of the Uber Tempe fatal collision in 2018. We used our notation and analysis method to produce a nuanced model showing different types of responsibility, such as moral attributability, accountability and liability. Additionally, we developed a picture of how responsibility issues (from multiple actors) during development led to operational shortfalls, ultimately contributing towards the accident.

The second example of an AI-based medical decision support tool, showed how the method can be used as a predictive analysis tool, proposing a number of additional risk reduction tasks, and illustrating the risks being mitigated by a clinician. To validate our work further, we need to demonstrate that our model and analyses have helped identify, and found means to alleviate, real safety responsibility issues. To a certain degree this has been accomplished via expert and peer review of our work. Additionally, we uncovered some missing roles and tasks, and also demonstrated how the burden of risk on the clinician may be increased.

For additional validation we could revisit the DCP responsibility model following an incident or accident. We could test whether the model and analysis had uncovered relevant issues and/or provided valuable evidence relating to responsibility. This is not practical, or ethical, to pro-actively pursue, but it remains a possibility if such an event occurs. 

As future work will develop the analysis methodology, including criteria for determining responsibility types such as moral accountability and liability. We will explore causal contribution in more depth, including severity and likelihood of impact, which also relates to accountability and liability. Inclusion of third-party actors with no specific role is also a future consideration. We additionally plan to develop a SAC framework incorporating the models. 

This work is potentially generalisable to other SCS, but our focus was on AI-SCS due to the known issues around responsibility.

%TC:ignore
\section*{Acknowledgements}
This work was supported by the Engineering and Physical Sciences Research Council (EP/W011239/1) and the Assuring Autonomy International Programme, a partnership between Lloyd’s Register Foundation and the University of York.
We would like to thank Paul Noordhof (Philosophy Department, University of York), Philip Morgan (York Law School, University of York), Tom Lawton (Bradford Teaching Hospitals) and Berk Ozturk (AAIP, University of York) for their invaluable insight and comments.

%TC:ignore
\bibliographystyle{ACM-Reference-Format}
\bibliography{newmain}
%TC:endignore

\end{document}